\documentclass[aps, amsmath,amssymb,prb,twocolumn,superscriptaddress]{revtex4-2}
\usepackage{graphicx,hyperref}
\usepackage[utf8]{inputenc}
\usepackage{amsmath,bm}
\usepackage{bbold}
\usepackage{color,soul}
\usepackage{xstring}
\usepackage{wasysym}
\usepackage{xspace}
\usepackage{newunicodechar}
\newunicodechar{：}{:}
\usepackage{amssymb}
\usepackage{time}
\usepackage{bbold}
\usepackage{subfigure}
\usepackage{multirow}
\usepackage{xspace}
\usepackage{url}
\usepackage{epstopdf}
\usepackage{cases}

\def\tg{\tilde{\Gamma}}
\def\<{\langle}
\def\>{\rangle}

\usepackage{ulem}

\begin{document}

\title{Periodicity-driven revision of the phase diagram of the generalized Baxter-Wu model with asymmetric complex couplings }
\author{\firstname{Yuting} \surname{Wang}}
\affiliation{\mbox{School of Physical Science and Technology, Beijing University of Posts and Telecommunications, Beijing 100876, China}}
\author{\firstname{Ye} \surname{Ling}}
\affiliation{\mbox{Department of Physics, Beijing Normal University, Beijing 100875, China }}
\author{\firstname{Haihong } \surname{Li}}
\affiliation{\mbox{School of Physical Science and Technology, Beijing University of Posts and Telecommunications, Beijing 100876, China}}
\author{\firstname{Yuhai} \surname{Liu}}
\email{yuhailiu@bupt.edu.cn}
\affiliation{\mbox{School of Physical Science and Technology, Beijing University of Posts and Telecommunications, Beijing 100876, China}}
\begin{abstract}
The conventional self-dual lines of the generalized Baxter-Wu (GBW) model with asymmetric complex couplings are known to be $\sinh(2K)=\pm\cos(2\phi)$, where $K$ and $\phi$ are the real and imaginary parts of the coupling. We demonstrate that these lines are incomplete: the periodicity of the partition function, encoded in the cosine factor of the bundled Boltzmann weight, generates additional self-dual lines $\sinh(2K)=\pm\sin(2\phi)$. Guided by the complete set of self-dual candidates, we perform Monte Carlo simulations using brute-force reweighting (Metropolis) and the Wang–Landau methods.
Simulations indicate that the self-dual lines at the partition-function minima $\phi_{\mathcal{Z}_{\min}}=(2n+1)\pi/8$ constitute a critical threshold. They are genuine critical boundaries for $|K| \ge \frac{1}{2}\operatorname{arsinh}(\cos(\pi/4)) \approx 0.32924$, while for smaller $|K|$ they are not.
At $\phi_{\mathcal{Z}_{\min}}$, the sign problem is most severe and finite-size scaling corrections are largest; the local peak observed below the phase boundary in the temperature scan is thus a finite-size artifact, not a genuine new phase.
We further clarify the capability and limitations of the average sign and its derivatives for detecting phase transitions. In particular, the negative peak of the average sign at $\phi_{\mathcal{Z}_{\min}}$ does not correspond to a genuine phase transition.
We also evaluate the Wang--Landau method, which, despite formally circumventing the sign problem, still faces the exponential barrier.
\end{abstract}

\maketitle

\section{Introduction}
Phase transitions and critical phenomena are among the central topics in condensed matter physics. Traditional phase transition theory has been constructed primarily for equilibrium systems, which are described by Hermitian Hamiltonians and therefore involve real parameters. The use of complex variables in statistical-mechanical models has a long history, beginning with the work of Lee and Yang on the Ising model in a complex magnetic field~\cite{PhysRev.87.410}. Later, researchers began to consider complex temperatures in the Ising model~\cite{Fisher_nature} or a complex number of states in the Potts model~\cite{Salas_transfer}. Meanwhile, open quantum systems with non-Hermitian Hamiltonians have also introduced complex couplings~\cite{yamamoto2026complexnonlinearsigmamodel, RevModPhys.88.035002, Non-Hermitian-Np2018}. More recently, complex couplings have been introduced into classical statistical models, including Potts~\cite{PhysRevLett.133.077101} and Baxter-Wu models~\cite{blote2017,signGBW}, where transfer-matrix mappings relate them to $(D-1)$-dimensional quantum systems~\cite{PhysRevLett.133.077101,signGBW}.

The Baxter-Wu model, describing three-spin interactions on a triangular lattice, was solved by Baxter and Wu in 1973~\cite{Baxter1973} and exhibits the four-state Potts critical behavior without logarithmic corrections~\cite{Baxter1973, Wu1982, Domany_BW-Potts}.
Deviations from this special point have been explored by extending model parameters~\cite{Youjin2010,blote2017,signGBW}. The up- and down-triangular interactions admit a self-dual generalization: asymmetric real couplings drive the transition to first order, while complex-conjugate couplings yield logarithmic corrections as in the four-state Potts model~\cite{blote2017,signGBW}.
In the complex-coupled model, pairing configurations via a $\pi$ rotation eliminates the imaginary part of the bundled weight, yielding a real weight containing a cosine factor~\cite{signGBW}. This cosine factor causes the sign problem in Monte Carlo simulations and also exhibits periodicity. In Ref.~\cite{signGBW}, we indirectly probed the critical behavior of the GBW model using universality by employing a reference system that shares the same symmetry with the original model and is defined by the absolute value of the bundled Boltzmann weight.
For the GBW model itself, apart from the self-dual lines near vanishing imaginary part that have been examined via transfer-matrix methods on a finite-width strip~\cite{blote2017}, but none of the self-dual lines have been systematically examined by unbiased Monte Carlo simulations. The influence of the cosine factor's periodicity on the phase diagram also remains unexplored.

The sign problem arises most acutely in Monte Carlo simulations of fermionic systems~\cite{PhysRevB.41.9301,santos2003introductionquantummontecarlo, PhysRevB.92.045110} or frustrated bosonic systems~\cite{10.1143/PTP.75.1254, HATANO1992246,PhysRevB.57.R3197}, as well as in certain complex-coupled systems~\cite{blote2017,signGBW}. The resulting cancellation causes an exponential growth in computational cost with system size, which is the exponential barrier that brute-force reweighting~\cite{PhysRevB.41.9301} reveals.
Despite decades of effort to resolve~\cite{Zi-Xiang2015} or alleviate~\cite{chang2023boosting, Karakuzu2023, Stefan2017, XiangT2016, PhysRevD.66.074507, PhysRevB.92.195126, PhysRevLett.126.216401, PhysRevD.86.074506, PhysRevC.63.034319} the sign problem, no universally applicable solution exists. The exponential barrier is caused by the exponential decay of the average sign with system size, whereas algebraic decay is non-universal, occurring only when the target and reference systems share the same ground-state energy and the degeneracy grows polynomially with size~\cite{PhysRevB.104.L241104,signbounds2022}.
Recently proposed sign-based phase transition probes~\cite{Mondaini2022} are reliable only when the reference free energy is flat; the modified sign scheme, though free of reference system interference, still cannot escape the exponential barrier~\cite{Mondaini2022, PhysRevB.110.125141,signGBW}. Ref.~\cite{signGBW} investigated sign-based phase transition probes using the known phase boundaries (self-dual lines) of the GBW model. Its phase diagram, however, did not account for the aforementioned periodicity of the cosine factor, nor had it been verified by unbiased Monte Carlo simulations. This motivates us to re-examine sign-based probes of phase transitions using this revised phase diagram.

A separate class of approaches circumvents the sign problem by directly sampling the density of states. The Wang–Landau algorithm~\cite{Wang-Landau2001, PhysRevE.64.056101, PhysRevLett.110.210603} uses the positive definite inverse density of states as its update probability, at least formally avoiding the sign problem. In quantum systems, Wang–Landau generalizes to the Linear Logarithmic Relaxation (LLR) algorithm~\cite{Gocksch1988, LLR2012, LLR2014}. LLR handles a $\mathbb{Z}_3$ spin model with complex action at polynomial complexity~\cite{LLR2014}, but for realistic fermionic systems, the exponential barrier may reappear~\cite{LLR2020}. Whether Wang–Landau/LLR can overcome the exponential barrier remains controversial. With its known transition line and amenability to the simpler Wang–Landau algorithm, the GBW model provides an ideal testing ground for this question.

To address these issues, we incorporate periodicity into the self-duality analysis, generating a discrete family of additional self-dual lines. Guided by this set, we perform large-scale Monte Carlo simulations using brute-force reweighting (Metropolis) and the Wang--Landau method. Based on the results, we
identify genuine critical boundaries for $|K| \geq \frac{1}{2}\operatorname{arsinh}(\cos(\pi/4))$, thereby obtaining a revised phase diagram.
At the partition-function minima $\phi_{\mathcal{Z}_{\min}}=(2n+1)\pi/8$, the sign problem is most severe and finite-size scaling corrections are largest; the local peak observed below the phase boundary in the temperature scan is thus a finite-size artifact, not a genuine new phase.
Based on this revised phase diagram, we re-examine sign-based phase transition probes and find that the minima of the derivative of the average sign correspond to the phase transition points of the reference system, while the minima of the average sign itself may correspond either to phase transitions or to non-critical regions. Finally, we find that the Wang–Landau method, similar to brute-force reweighting, likewise fails to circumvent the exponential barrier in practice. These results establish a periodicity-corrected self-duality framework, clarify the detailed limitations of sign-based phase transition probes, and assess the exponential barrier of the Wang–Landau method.

The paper is organized as follows. Sec.~\ref{sec:model} presents the GBW model and its self-duality analysis. Sec.~\ref{sec:method} describes the Monte Carlo methods. Sec.~\ref{sec:results} discusses the simulation results. Sec.~\ref{sec:conclusions} presents the conclusions and discussion.

\begin{figure}[htb]
  \includegraphics[width=0.45\textwidth]{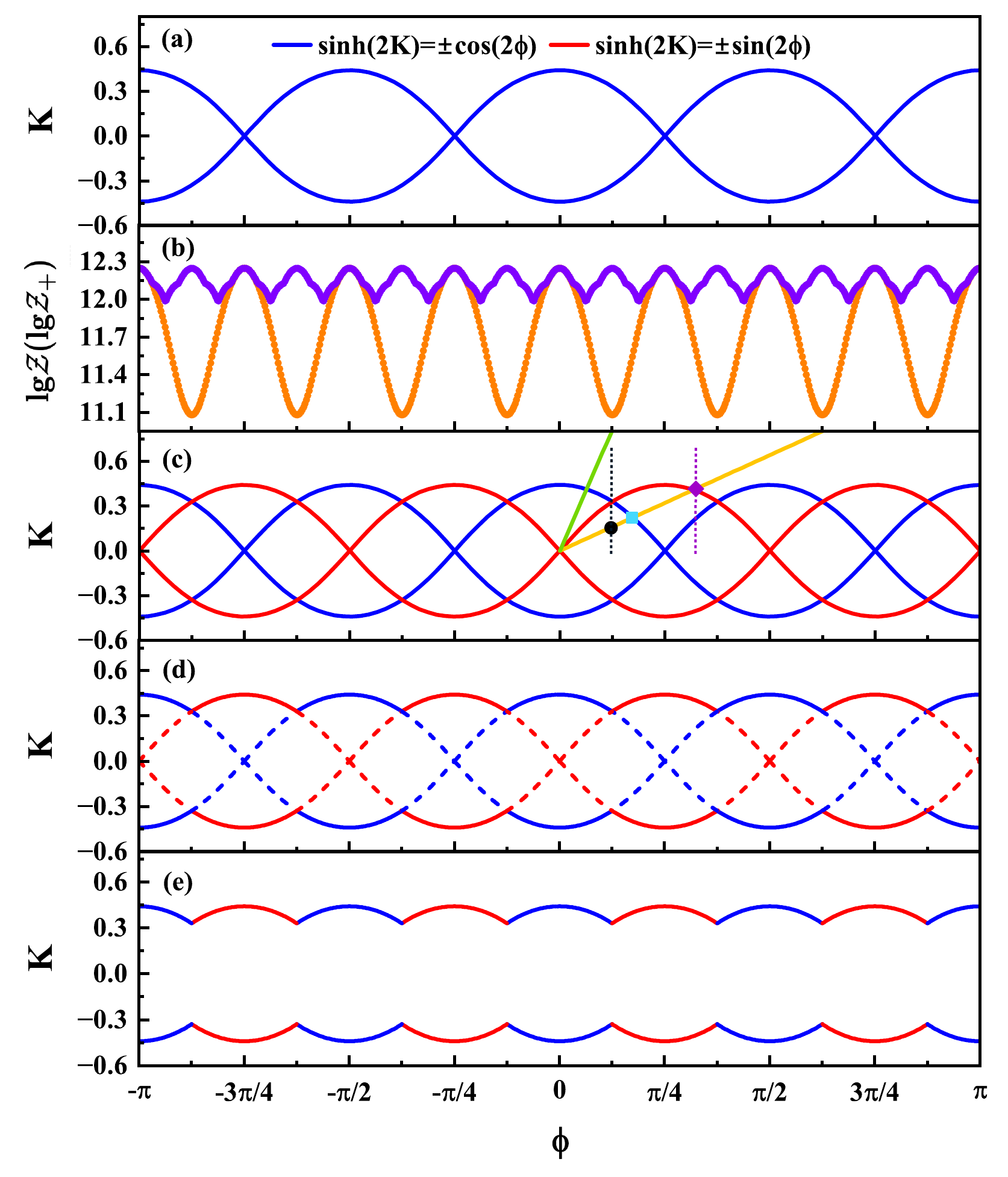}
  \caption{\label{fig:phase_diagram}
 Schematic of phase-diagram analysis. (a) Conventional self-dual phase diagram of the GBW model from Eq.~(\ref{Eq:self_dual_A}). (b) For $K=0.3$, the complex-coupled GBW model (orange) has a partition-function period of $\pi/4$, while the reference system (purple) has a period of $\pi/8$. (c) Complete phase diagram after incorporating periodicity: newly identified self-dual lines (SDB) in red, conventional ones (SDA) in blue. (d) Monte Carlo validation: solid segments are genuine critical boundaries; dashed segments are excluded. (e) Final revised phase diagram of the complex-coupled GBW model.
}
\end{figure}

\section{Model and self-duality analysis}
\label{sec:model}
We consider the generalized Baxter-Wu (GBW) model with asymmetric three-spin interactions in the up- and down triangles described by the Hamiltonian:
\begin{equation}
\label{Eq:Ham}
\beta H =- K_{\rm up}\sum_\triangle \sigma _{i}\sigma _{j}\sigma_{k}-K_{\rm down} \sum_{\bigtriangledown } \sigma _{l}\sigma _{m}\sigma_{n},
\end{equation}
where $\sigma_i = \pm 1$ is the Ising spin at site $i$, $\beta = 1/(k_{\rm B}T)$ is the inverse temperature, and the sums run over all triangles, $K_{\rm up}$ and $K_{\rm down}$ are the couplings on up- and down-triangles, respectively. The model is self-dual on lines~\cite{Youjin2010}:
$\sinh (2K_{\text{up}}) \sinh (2K_{\text{down}}) = 1$.
For the mutually complex-conjugate couplings $K_{\rm up}=K+i\phi$, $K_{\rm down}=K-i\phi$, it reduces to~\cite{signGBW}:
\begin{equation}
\label{Eq:self_dual_A}
\sinh (2K) = \pm \cos (2\phi),
\end{equation}
as shown in Fig.~\ref{fig:phase_diagram}(a).
These two self-dual lines separate a disordered phase from two ordered phases.
On the triangular lattice, there are three sublattices A, B, and C. For $K>0$, the four ground states are $(\sigma _A,\sigma _B,\sigma_C)=(+,+,+)$, $(+,-,-)$, $(-,+,-)$, and $(-,-,+)$; for $K<0$, they are obtained by flipping all spins.

\begin{figure}[htb]
  \includegraphics[width=0.4\textwidth]{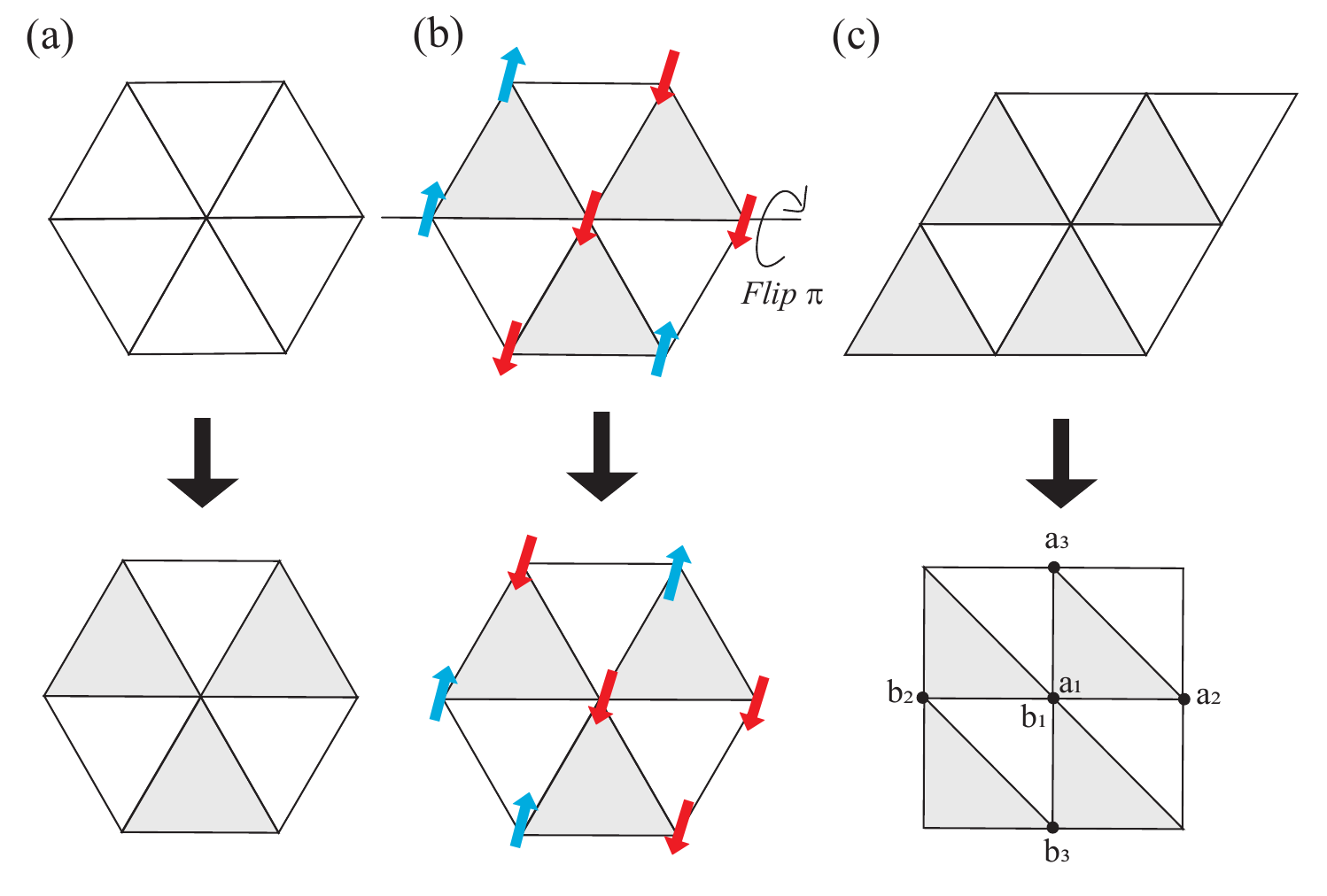}
  \caption{\label{fig:triangular_lattice}
  Schematic diagram of the lattice. (a) Baxter-Wu lattice with $C_6$ symmetry (top) and asymmetric couplings reducing it to $C_3$ (bottom). (b) A $\pi$ rotation pairs two configurations, removing the imaginary part of the bundled weight~\cite{signGBW}. (c) Triangular lattice mapped onto a square lattice with diagonal bonds.}
\end{figure}

In the vicinity of the $K_{\rm up}=K_{\rm down}$ limit, both asymmetric real couplings and mutually complex-conjugate couplings break the lattice rotational symmetry from $C_6$ to $C_3$ [Fig.~\ref{fig:triangular_lattice}(a)] without affecting the spontaneous breaking of $S_4$ symmetry that accompanies the phase transition. However, these couplings qualitatively alter the nature of the transition in different ways. The real couplings drive it into the first-order regime. In contrast, the complex-conjugate couplings cause the system to deviate from the four-state Potts fixed point, leading to logarithmic corrections; this scenario can be mapped onto a one-dimensional quantum model via a transfer-matrix representation~\cite{blote2017, signGBW}.

For the complex-coupling case, the Boltzmann weight of a given configuration $\Gamma\equiv(\sigma_1,\sigma_2,\ldots,\sigma_N)$ is complex-valued, as shown in Eq.~(\ref{Eq:Boltzmann}). As described in Ref.~\cite{signGBW} and Fig.~\ref{fig:triangular_lattice}(b), pairing $\Gamma$ with its $\pi$-rotated counterpart $\Gamma'$ about a lattice direction yields the weight of the bundled configuration $\tilde{\Gamma}$:
\begin{equation}
\label{Eq:Wbindingconf}
	\begin{split}
		W(\tilde{\Gamma}, T) &\equiv  W(\Gamma, T)+W(\Gamma', T)\\
		&=2e^{KH_{K}(\Gamma) }\cos\left (\phi H_{i\phi}(\Gamma)\right ) .\\
   \end{split}
\end{equation}
Here the temperature $T$ is absorbed into $K$ and $\phi$, the $H_K$ and $H_{i\phi}$ are derived from the partial derivatives of the Hamiltonian, as shown in Eqs.~(\ref{eq:app_HK}) and (\ref{eq:app_Hiph}) of the Appendix~\ref{app:partition}: $H_K$ corresponds to the sum over all up- and down-triangles, while $H_{i\phi}$ corresponds to the difference between the up- and down-triangle sums. The imaginary part of the bundled weight is thereby eliminated, leaving a multiplicative cosine factor that oscillates in sign. This oscillation causes the sign problem in Monte Carlo simulations; the cosine factor also possesses periodicity. Then the partition function is:
\begin{equation}
	\mathcal{Z}(T)= \sum_{\Gamma} W(\Gamma, T)=\frac{1}{2} \sum_{\tilde{\Gamma}} W(\tilde{\Gamma}, T),
    \label{Zbw}
\end{equation}

As shown in Ref.~\cite{signGBW}, the bundled weight is periodic in $\phi$ with period $\pi/4$, because a single spin flip changes $H_{i\phi}(\Gamma)$ [corresponding to $A(\Gamma)-B(\Gamma)$ in Ref.~\cite{signGBW}] by multiples of 8 and the cosine has period $2\pi$. We can further infer that the partition function in Eq.~(\ref{Zbw}) exhibits a period of $\pi/4$ in $\phi$. This periodicity is also verified by exact enumeration for a $6\times6$ system, as shown by the orange curve in Fig.~\ref{fig:phase_diagram}(b).
Remarkably, the $\pi/4$ periodicity of the partition function is in sharp contradiction with the conventional self-dual phase diagram in Fig.~\ref{fig:phase_diagram}(a), which exhibits a $\pi/2$ period in $\phi$. This discrepancy motivates us to re-examine previous self-duality analyses~\cite{Youjin2010,blote2017}.

We follow the approach of Ref.~\cite{Youjin2010} and map the model onto a two-state Potts model on a square lattice with extra diagonal bonds, as shown in Fig.~\ref{fig:triangular_lattice}(c). The corresponding partition function is then written as:
\begin{equation}
\begin{aligned}
Z(v_1, v_2) = \sum_{\{\sigma_r\}} \prod_r \left[ 1 + v_1 \delta_2 \left( \sum_{i=1}^n \sigma_{r+a_i} \right) \right] \\\left[ 1 + v_2 \delta_2 \left( \sum_{j=1}^m \sigma_{r+b_j} \right) \right]
\end{aligned}
\label{eq:origin partition function}
\end{equation}
where $\mathbf r$ runs over all lattice sites, and the displacement vectors are chosen as:
\[
\mathbf a_1 = \mathbf b_1 = (0,0),\;
\mathbf a_2 = -\mathbf b_2 = (1,0),\;
\mathbf a_3 = -\mathbf b_3 = (0,1).
\]
The function $\delta_q(x)$ equals $1$ when $x\bmod q=0$ and $0$ otherwise. In this Potts formulation, the three-spin product maps to $0$ or $1$, and the couplings are doubled accordingly. When the aforementioned periodicity is taken into account, $v_1$ and $v_2$ satisfy:
\begin{equation}
\left\{
\begin{aligned}
v_1 &= \exp\Bigl(2\bigl[K + i(\phi + n\pi/4)\bigr]\Bigr) - 1, \\
v_2 &= \exp\Bigl(2\bigl[K - i(\phi + n\pi/4)\bigr]\Bigr) - 1.
\end{aligned}
\right.
\label{v1v2}
\end{equation}
The duality transformation of Eq.~(\ref{eq:origin partition function}) yields the self-duality relation $v_1 v_2 = 2$. Substituting Eq.~(\ref{v1v2}) and simplifying yields the complete self-dual line as $\sinh(2K) = \cos(2\phi + n\pi/2)$, which can be further divided into two families:
\begin{subnumcases}{\sinh(2K) =}
\label{Eq:SDA}
\pm \cos(2\phi), & $n$ even, \label{Eq:self_dual_line1} \\[4pt]
\label{Eq:SDB}
\pm \sin(2\phi), & $n$ odd. \label{Eq:self_dual_line2}
\end{subnumcases}
The former corresponds to the conventional self-dual line in Fig.~\ref{fig:phase_diagram}(a), while the latter corresponds to the new family of dual lines discovered from the periodicity of the partition function, with the complete set shown in Fig.~\ref{fig:phase_diagram}(c).
Since self-duality alone does not guarantee that every segment of a self-dual line is a genuine phase boundary, efficient Monte Carlo simulations are required to examine the complete set.
\section{Monte Carlo methods}
\label{sec:method}
The GBW model with asymmetric complex coupling suffers from a sign problem originating from the cosine factor in Eq.~(\ref{Eq:Wbindingconf}). We employ two Monte Carlo methods: brute-force reweighting~\cite{PhysRevB.41.9301} combined with the Metropolis algorithm and the Wang--Landau (WL) algorithm~\cite{Wang-Landau2001, PhysRevE.64.056101, PhysRevLett.110.210603}. Both formally avoid direct sampling with the signed weight.
\subsection{Brute-force reweighting technique}
In brute-force reweighting, we introduce a reference model with a partition function:
\begin{equation}
    \mathcal{Z}_+(T) = \frac{1}{2} \sum_{\tilde{\Gamma}} |W(\tilde{\Gamma}, T)|.
    \label{Zbw+}
\end{equation}
The expectation value of an observable $O$ is:
\begin{equation}
\begin{aligned}
\label{RWo}
\langle O \rangle &= \frac{\sum_{\tg} O(\tg) \, W(\tg, T)}{\sum_{\tg} W(\tg, T)} \\
%&= \frac{\sum_{\tg} O(\tg) \, S(\tg) |W(\tg, T)|}{\sum_{\tg} S(\tg)|W(\tg, T)|} \\
&= \frac{\sum_{\tg} O(\tg) \, S(\tg) \, p^{+}(\tg)}{\sum_{\tg} S(\tg) \, p^{+}(\tg)} \\
&= \frac{\langle O \cdot S \rangle_{+}}{\langle S \rangle_{+}}.
\end{aligned}
\end{equation}
Here, $p^{+}(\tg) = \frac{|W(\tg, T)|}{\sum_{\tg} |W(\tg, T)|}$ is the probability distribution of the reference model, $S(\tg) = \frac{W(\tg, T)}{|W(\tg, T)|}$ is the sign of a configuration, and $\langle \cdot \rangle_{+}$ denotes the average with respect to $p^{+}$. The average sign $\langle S \rangle_+$ is expressed as:
\begin{equation}
\begin{aligned}
\label{Eq:average_sign1}
\left \langle S \right \rangle_+ =\frac{\sum_{\tg} \left| W(\tg, T )\right | S(\tg)}
{\sum_{\tg}\left |W(\tg, T)\right|}  =\frac{{\cal Z}(T)}{{\cal Z}_+(T)}
 =e^{-\beta \Delta F}.
\end{aligned}
\end{equation}
Here, the free energy $F = -k_B T \ln \mathcal{Z}$, $\Delta F=N \Delta \mathrm{f} $ is the free-energy difference between the model Eq.~(\ref{Zbw}) and its reference system Eq.~(\ref{Zbw+}), then $\left \langle S \right \rangle_+\propto e^{-\beta N}$, which decays exponentially with system size. To achieve a given accuracy, the computation time scales as,
\begin{equation}
\label{Eq:Msteps}
T_{Reweighting} \propto 1 / \left \langle S \right \rangle_+^2 \propto e^{2\beta N},
\end{equation}
which is the so-called exponential barrier.
The exponential decay of the average sign is a characteristic feature of most sign-problematic systems, while algebraic decay has been observed only in a few specific models~\cite{PhysRevB.104.L241104,signbounds2022}.
\subsection{Wang-Landau algorithm}
In general, the partition function can be written as:
\begin{equation}
	\mathcal{Z}(T)= \sum_{\Gamma} W(\Gamma, T)=\sum_{E} g(E)W(E, T),
    \label{ZPF}
\end{equation}
which replaces the exponentially growing sum over configurations with an algebraically growing sum over energy levels. The WL algorithm~\cite{Wang-Landau2001, PhysRevE.64.056101} bypasses the Boltzmann weight and directly estimates the density of states $g(E)$.
The algorithm performs a random walk in energy space with acceptance probability:
\begin{equation}
P_{\rm accept} = \min\!\left[\frac{g(E_1)}{g(E_2)},\ 1\right],
\label{eq:wl_accept_en}
\end{equation}
and updates $g(E) \to g(E) \times f$ upon each visit while accumulating a histogram $H(E)$. When $H(E)$ becomes sufficiently flat, $f$ is reduced as $f \to \sqrt{f}$ and $H(E)$ is reset. This process is repeated until $f$ falls below a preset threshold, yielding a high-precision estimate of $g(E)$, from which thermodynamic quantities such as the free energy and specific heat can be computed.

For the complex coupled GBW model, in the bundled-configuration scheme described in Sec.~\ref{sec:model}, the energy for the bundled configuration $\tilde{\Gamma}$ is $\beta E(\tg)= -KH_{K} + \phi H_{i\phi}\tan(\phi H_{i\phi})\bigr.$ (See Eq.~(\ref{eq:E_bundled}) for details.)
A difficulty arises, however, in that the bundled weight in Eq.~(\ref{Eq:Wbindingconf}) cannot be expressed as a function of the energy as required by Eq.~(\ref{ZPF}). Nevertheless, since both quantities are functions of $H_{K}$ and $H_{i\phi}$, we perform the WL simulation in the parameter space defined by $(H_{K}, H_{i\phi})$. The partition function in Eq.~(\ref{ZPF}) is then expressed as:
\begin{equation}
\label{ZPFg}
\mathcal{Z}(T) = \sum_{(H_{K},H_{i\phi})} g(H_{K},H_{i\phi}) W(H_{K}, H_{i\phi}),
\end{equation}
Here, $g(H_{K},H_{i\phi})$ denotes the density of states.
The number of $(H_K, H_{i\phi})$ combinations, i.e., the number of energy levels, is on the order of $L^4$, as Eq~(\ref{eq:NHkHiphi}) detailed in the Appendix~\ref{app:partition}.
For comparison, in the pure Baxter-Wu model, the number of energy levels scales only as $L^2$. The much faster growth in the present case leads directly to slow convergence, which in turn limits our WL simulations.

To accelerate convergence, we divide the parameter space into overlapping sub-windows following Ref.~\cite{PhysRevLett.110.210603}, with about $75\%$ overlap. An independent WL walker is assigned to each subwindow with the standard iteration scheme: flatness threshold $\alpha = 0.98$, reduction $f \to \sqrt{f}$, and termination at $\ln f_{\mathrm{min}} = 10^{-7}$. After convergence, the piecewise densities of states are joined into a global $g(H_{K}, H_{i\phi})$. Although these efforts only allow us to reach $L \le 15$, this already far exceeds the $L=6$ accessible by full enumeration.% We have thus obtained the global density of states.

The WL algorithm is well suited for this study. The acceptance probability relies solely on the positive-definite density of states $g(H_K, H_{i\phi})$, and the equal-visitation sampling reduces critical slowing down. Directly computing the free energy and its derivatives enables exploration of novel phases without an order-parameter prior, while a single run provides all thermodynamic observables at any temperature.
\section{Numerical results}
\label{sec:results}
\subsection{Monte Carlo verification of the complete self-dual phase diagram}
\label{MC_verification}
To identify which self-dual lines are genuine phase boundaries and the nature of the transitions, we perform Monte Carlo simulations on the full self-dual phase diagram shown in Fig.~\ref{fig:phase_diagram}(c). For clarity, we denote the conventional self-dual lines reported in Refs.~\cite{blote2017,signGBW} as self-dual line A (SDA) and the new ones found in this work as self-dual line B (SDB).

\begin{figure}[htb]
	\includegraphics[width=0.45\textwidth]
	{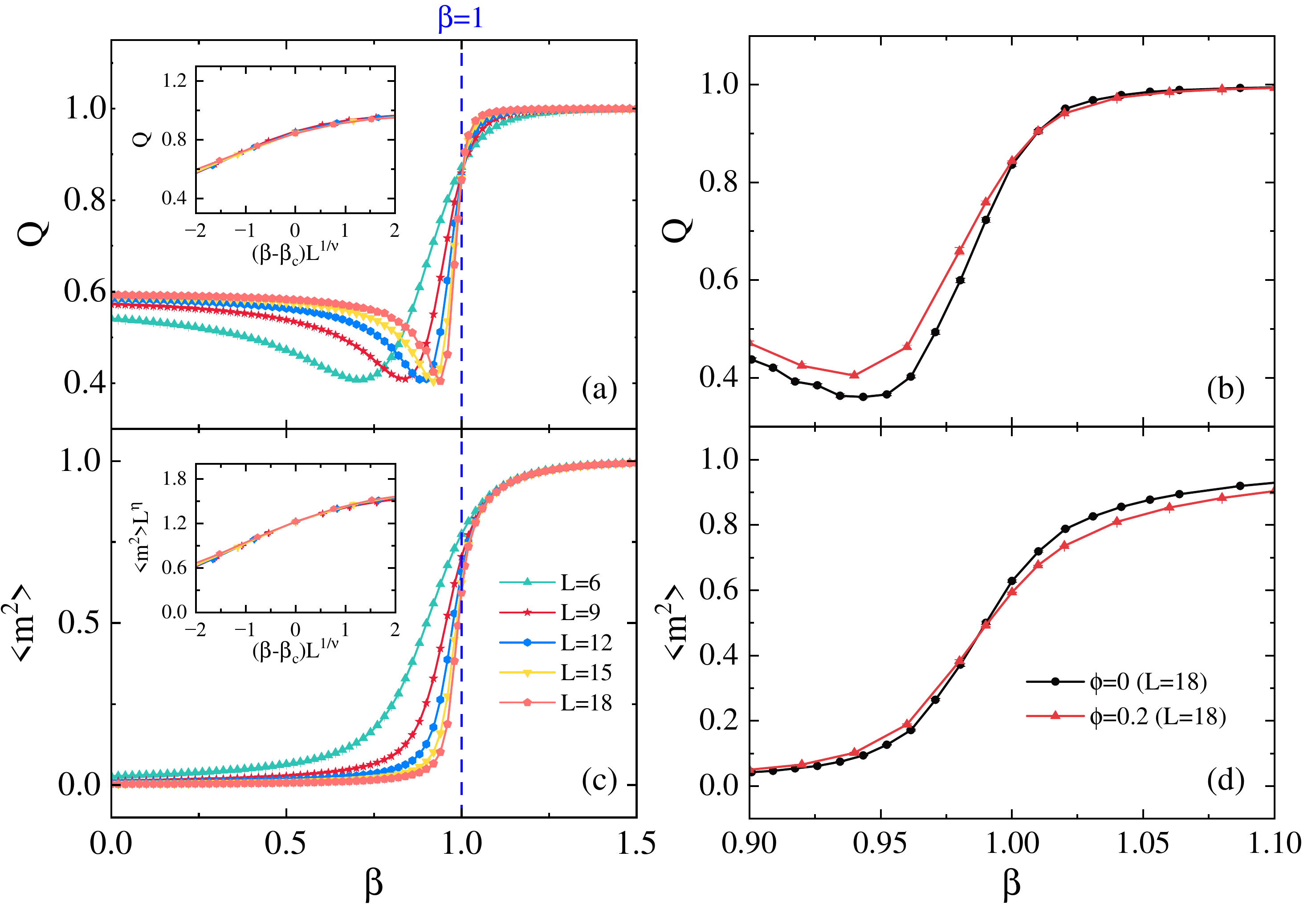}
	\caption{(a),(c) Binder ratio $Q$ and squared order parameter $\langle m^2\rangle$ as functions of inverse temperature $\beta$ along the temperature path $(K,\phi)=\beta(0.41222,0.2)$, which intersects the SDA line at $\beta=1$. The insets in (a) and (c) show the data collapse with $\eta=1/4$ and $\nu=2/3$.  (b),(d) Comparison of $\langle m^2\rangle$ and $Q$ at $L=18$ between the complex-coupled GBW model ($\phi=0.2$) and the pure Baxter--Wu model ($\phi=0$), showing milder temperature dependence in the former.}
	\label{fig:beta-1}
\end{figure}

To study the phase transition from the fourfold-degenerate ground states to the high-temperature disordered phase, the order parameter is defined in analogy with the q-state Potts model:
\begin{equation}
\label{Eq:order_parameter}
	m^2=\frac{1}{q-1}\sum_{i=1}^{q-1}\sum_{j=i+1}^q(\rho_i-\rho_j)^2
\end{equation}
where~$q=4$~and  $i,j$ run over the four types of triangles: $(+,+,+)$, $(+,-,-)$, $(-,+,-)$, and $(-,-,+)$, associated with the corresponding ground states, and $\rho_i$ is the density of triangles in the $i$-th ground state.
The fourth-order Binder ratio of $m$ is defined as:
\begin{equation}
\label{Eq:Binder_ratio}
	Q(T, L) =\frac{\langle m^2\rangle^2}{\langle m^4\rangle},
\end{equation}
which is renormalization invariant at the critical point. Using the brute-force reweighting technique described in Sec.~\ref{sec:method}, we perform Metropolis Monte Carlo simulations of the reference system in Eq.~(\ref{Zbw+}) to compute the squared order parameter and the Binder ratio, along with their statistical errors.

We select the two temperature paths indicated in Fig.~\ref{fig:phase_diagram}(c). The green line, parameterized by $(K,\phi)=\beta(0.41222,0.2)$ with $\beta$ the reduced inverse temperature, intersects SDA at $\beta=1$. The yellow line, parameterized by $(K,\phi)=\beta(0.40788,1.0)$, crosses SDB at $\beta=1$ and SDA at $\beta=0.54624$.

A transfer-matrix study~\cite{blote2017} showed that SDA is a genuine phase transition line for $\phi\in(0,0.32]$, belonging to the four-state Potts universality class with logarithmic corrections.
Here we re-examine this result along the temperature path $(K,\phi)=\beta(0.41222,0.2)$. As shown in Fig.~\ref{fig:beta-1}(a) and (c), the Binder ratio $Q$ exhibits a clear crossing at the intersection of the temperature path with the SDA line ($\beta=1$). The squared order parameter $\langle m^2\rangle$ as a function of inverse temperature also shows a progressively steeper rise with increasing system size at $\beta=1$. As shown in Fig.~\ref{fig:beta-1}(b) and (d), comparing the complex-coupled GBW model at $\phi=0.2$ with the pure Baxter--Wu model for $L=18$, we find that both quantities exhibit a milder temperature dependence in the complex-coupled case, suggesting a continuous transition. To identify the universality class of this transition, we apply a data-collapse procedure based on the finite-size scaling ansatz~\cite{Luck1985}. In two dimensions ($d=2$), this ansatz reduces to:
\begin{align}
m^2(L,g) &= L^{-\eta} f\big((g - g_c)L^{1/\nu}\big), \label{eq:m2}\\
Q(L,g) &= f\big((g - g_c)L^{1/\nu}\big), \label{eq:Q}
\end{align}
where $L$ is the linear size, $\nu$ and $\eta$ are the correlation-length exponent and anomalous dimension, respectively, and $g$ denotes the tuning parameter ($\beta$ or $K$ in this work). As shown in the insets of Fig.~\ref{fig:beta-1}(a) and (c), the data for different system sizes collapse well onto a single curve with $\eta=1/4$ and $\nu=2/3$, confirming that the phase transition belongs to the four-state Potts universality class, in agreement with the prediction of Ref.~\cite{blote2017}. Owing to the sign problem, our calculations are restricted to relatively small system sizes ($L\le18$), which precludes examining the logarithmic corrections.

\begin{figure}[htb]
	\includegraphics[width=0.45\textwidth]
	{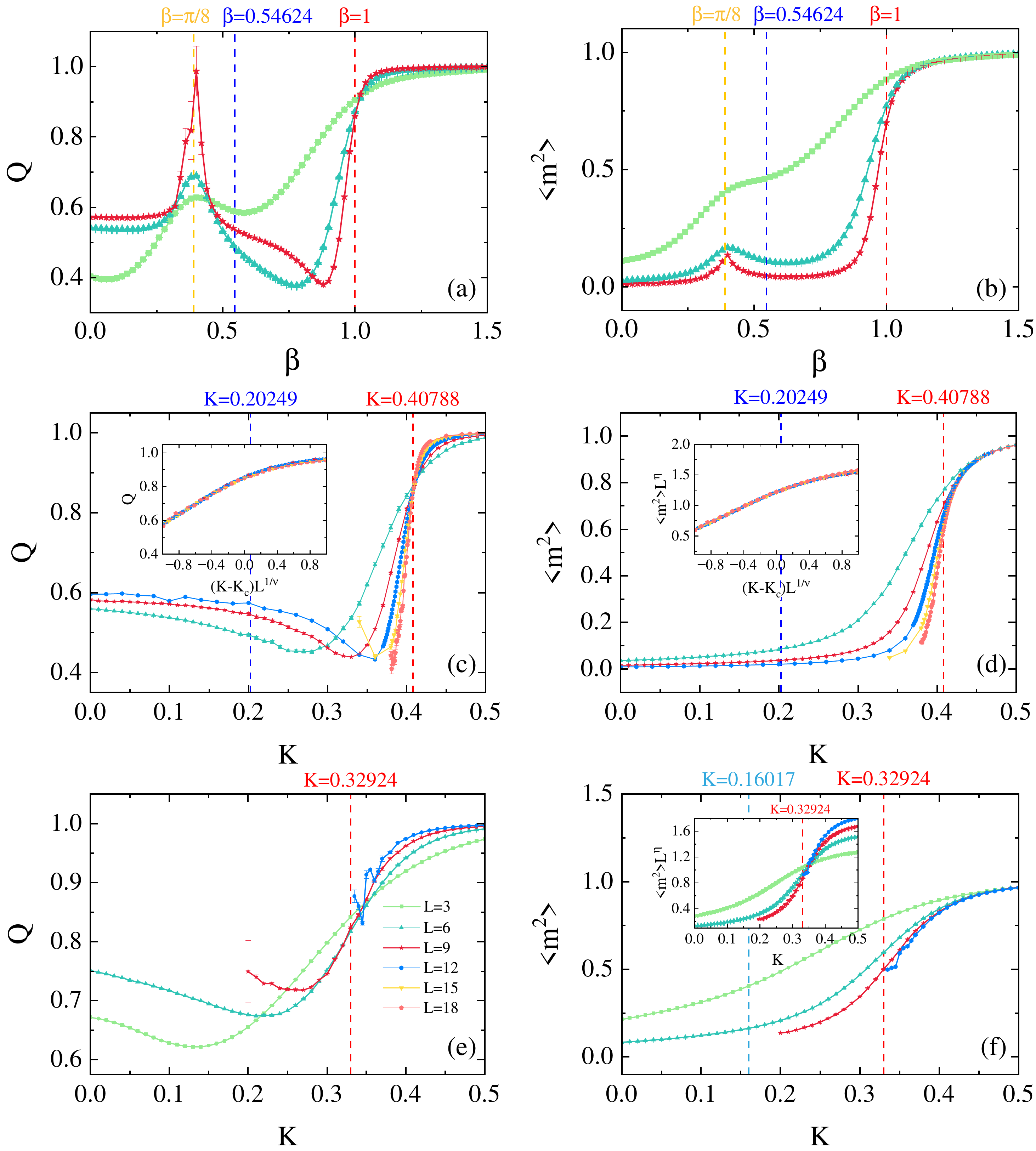}
	\caption {(a),(b) $Q$ and $\langle m^2\rangle$ vs $\beta$ along $(K,\phi)=\beta(0.40788,1.0)$. (c),(e) $Q$ and (d),(f) $\langle m^2\rangle$ vs $K$ at $\phi=1$ and $\phi=\pi/8$. For $\phi=1$ [(c),(d)], a transition appears only on the solid segments of the self-dual lines in Fig.~\ref{fig:phase_diagram}(d); data collapse gives $\nu=2/3,\eta=1/4$, consistent with the 4-state Potts class. For $\phi=\pi/8$ [(e),(f)], $Q$ crossings appear near the intersection of SDA and SDB, but the $L=12$ data are too noisy; in contrast, the scaled quantity $\langle m^2\rangle L^{\eta}$ clearly locates $K_c$.}
	\label{fig:longitudinal}
\end{figure}

We next examine the second temperature path, $(K,\phi)=\beta(0.40788,1.0)$, which crosses both SDA and SDB [Fig.~\ref{fig:longitudinal}(a),(b)]. The Binder ratio $Q$ shows a clear crossing near SDB ($\beta=1$) but none near SDA ($\beta=0.54624$); the squared order parameter $\langle m^2\rangle$ rises steeply with system size at SDB while remaining flat at SDA. These observations establish that SDB is a genuine phase boundary and that SDA is not at this parameter value. We further infer that both SDA and SDB contain only segments that correspond to true phase boundaries, as shown in Fig.~\ref{fig:phase_diagram}(d): the solid lines indicate genuine critical boundaries, while the dashed segments do not.

We fix $\phi=1$ and measure $Q$ and $\langle m^2\rangle$ as functions of $K$ [Fig.~\ref{fig:longitudinal}(c),(d)] to examine the universality class of the SDB line while avoiding the complexity of the $\phi$-dependent sign problem, thereby further verifying our revised phase diagram. A single transition point is observed on the solid segment of the self-dual lines, with no signal on the dashed segments of Fig.~\ref{fig:phase_diagram}(d). Data collapse with $\nu=2/3$ and $\eta=1/4$ (insets of Fig.~\ref{fig:longitudinal}(c),(d)) confirms the four-state Potts universality class.

As shown in Fig.~\ref{fig:longitudinal}(a), a surprising signal emerges as a peak at $\beta = \pi/8$ that grows with system size. Since $(K,\phi)=\beta(0.40788,1.0)$, $\beta = \pi/8$ corresponds to $\phi = \pi/8$,  which is the minimum of the partition function in Fig.~\ref{fig:phase_diagram}(b) and the point where the sign problem is most severe~\cite{signGBW}.
A similar signal also appears in the plot of the squared order parameter in Fig.~\ref{fig:longitudinal}(b). The peak at $\beta = \pi/8$ decreases rapidly with increasing system size; however, since the sign problem is most severe at this point, we are unable to simulate larger system sizes to perform a finite-size extrapolation and thus cannot rule out long-range order at this point.

\begin{figure}[htb]
	\includegraphics[width=0.45\textwidth]{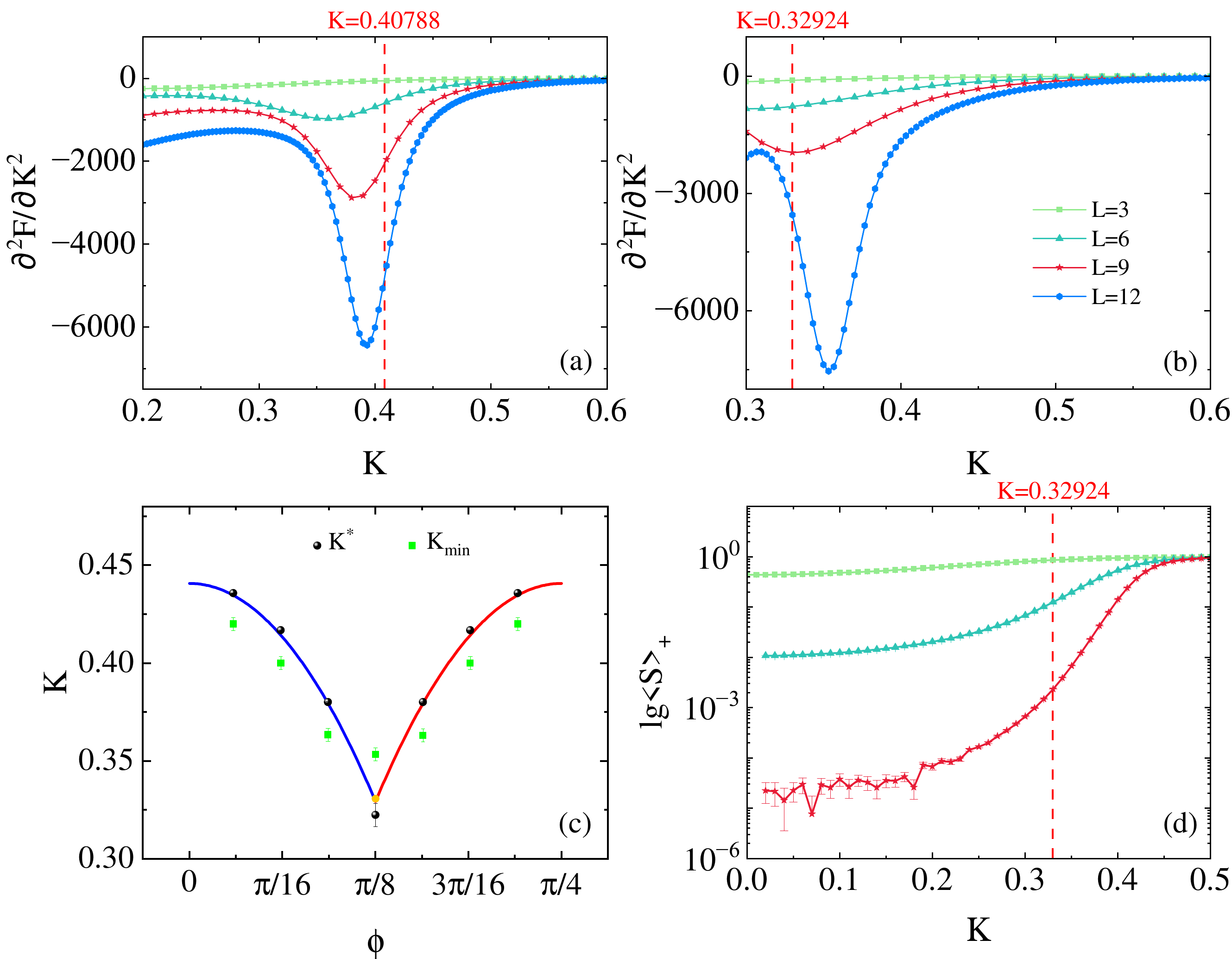}
	\caption{(a),(b) Second derivative of the free energy, $\partial^2 F/\partial K^2$, from the WL algorithm at $\phi=1$ (a) and at $\phi=\pi/8$ (b). In (a), the negative peak shifts toward the transition point with $L$; in (b), where the sign problem is most severe and logarithmic corrections are largest, the peak converges much more slowly. (c) $K_{\rm min}$ denotes the location of the minimum of $\partial^2 F/\partial K^2$ at $L=12$, while $K^*$ corresponds to the Binder-ratio crossings for $(L_1,L_2)=(12,18)$, $(9,12)$ (twice), and $(6,9)$, ordered from left to right in the interval $(0,\pi/8]$; both sets agree with the inferred transition line. (d) $\lg\langle S\rangle_+$ at $\phi=\pi/8$ as a function of $K$.
    }
	\label{fig:WL-longitudinal}
\end{figure}

To clarify the behavior near this point, we also investigate $Q$ and $\langle m^2\rangle$ along the $\phi=\pi/8$ axis [Fig.~\ref{fig:longitudinal}(e),(f)]. A clear crossing of $Q$ appears near the SDA--SDB intersection at $(\pi/8, \frac{1}{2}\operatorname{arsinh}(\cos(\pi/4)))$, but the $L=12$ data are noisy, making it unclear whether the crossing converges to the intersection in the thermodynamic limit. This is because $Q$ depends on $\langle m^4\rangle$, which is more sensitive to fluctuations than $\langle m^2\rangle$; the sign problem further amplifies this via cancellations in the Monte Carlo weights. Reliable estimates of $Q$ thus require longer simulation time than $\langle m^2\rangle$, especially at small $K$, where the sign problem worsens [Fig.~\ref{fig:WL-longitudinal}(d)]. By contrast, $\langle m^2\rangle$, being a lower-order moment, converges faster with smaller statistical errors. We therefore apply the finite-size scaling ansatz Eq.~(\ref{eq:m2}) to $\langle m^2\rangle$. The scaled quantity $\langle m^2\rangle L^{\eta}$ exhibits clear intersections across system sizes with $\eta=1/4$, yielding a crossing point that approaches the self-dual intersection $(\pi/8, \frac{1}{2}\operatorname{arsinh}(\cos(\pi/4)))$. This confirms that the transition coincides with the self-dual line.

For the anomalous parameter point at $\phi = \pi/8$ in Fig.~\ref{fig:longitudinal}(a) and (b), corresponding to the black dot in the phase diagram Fig.~\ref{fig:phase_diagram}(c). Solving the path equation for this point gives $K\approx0.16017$. However, neither $\langle m^2\rangle$ nor $Q$ shows an anomalous signal at this value. Nevertheless, the sign problem [Fig.~\ref{fig:WL-longitudinal}(d)] restricts our simulations to very small system sizes ($L \le 6$), preventing a definitive exclusion of finite-size effects. Therefore, our preliminary conclusion is that the ferromagnetic-paramagnetic phase transition still occurs at the self-dual point.

So far, our analysis of the phase diagram has relied solely on the order parameter defined in Eq.~(\ref{Eq:order_parameter}), which does not rule out the presence of new phases or phase transitions. To further explore the phase diagram, we perform simulations using the WL algorithm. From the resulting density of states $g(H_K, H_{i\phi})$, we extract the partition function Eq.~(\ref{ZPFg}) and the free energy $F = -T \ln \mathcal{Z}$, from which we compute the second derivative of the free energy with respect to $K$:

\begin{equation}
\frac{\partial^2 F}{\partial K^2} = -\frac{T}{\mathcal{Z}} \frac{\partial^2 \mathcal{Z}}{\partial K^2} + \frac{T}{\mathcal{Z}^2} \left(\frac{\partial \mathcal{Z}}{\partial K}\right)^2,
\end{equation}
where $\frac{\partial\mathcal{Z}}{\partial K}$ and $\frac{\partial^2\mathcal{Z}}{\partial K^2}$are given by Eqs.~(\ref{eq:app_Zp_derivK1}) and (\ref{eq:app_Zp_derivK2}) in the Appendix~\ref{app:partition} .

As shown in Fig.~\ref{fig:WL-longitudinal}(a), at $\phi=1.0$, the second derivative of the free energy exhibits a single negative peak whose position shifts toward the self-dual line $|K| \geq \frac{1}{2}\operatorname{arsinh}(\cos(\pi/4))$ with increasing system size. No signal is detected on the self-dual line with $|K| < \frac{1}{2}\operatorname{arsinh}(\cos(\pi/4))$. This is consistent with the genuine phase boundary identified through the order-parameter analysis. For $\phi=\pi/8$ as shown in Fig.~\ref{fig:WL-longitudinal}(b), the observed transition signal lies slightly above the self-dual point. This is because $\phi=\pi/8$ corresponds to the point where logarithmic corrections are most severe and finite-size effects are strongest, leading to the slowest convergence to the self-dual line with system size. The sign problem in Fig.~\ref{fig:WL-longitudinal}(d) prevents us from accessing smaller $K$ values, limiting the available range for finite-size scaling.

We focus on the interval $[0,\pi/4]$ as a representative region in which to verify the phase boundary; the remaining intervals follow by periodicity. The sign problem prevents a reliable finite-size extrapolation to the thermodynamic limit. Nevertheless, within this interval, the Binder-ratio crossings and the minima of the second derivative of the free energy (at $L=12$) are consistent with the conjectured phase boundary Fig.~\ref{fig:WL-longitudinal}(c).

Taken together, our data confirm that, at the minima of the partition function $\phi_{\mathcal{Z}_{\min}} = (2n+1)\pi/8$, the self-dual lines define a critical threshold: for $|K| \geq \frac{1}{2}\operatorname{arsinh}(\cos(\pi/4))$ they are genuine critical boundaries, while for $|K| < \frac{1}{2}\operatorname{arsinh}(\cos(\pi/4))$ they are not as shown in Fig.~\ref{fig:phase_diagram}(e). Our Monte Carlo simulations tentatively confirm, through data collapse, that the phase transition belongs to the four-state Potts universality class.
At $\phi_{\mathcal{Z}_{\min}}=(2n+1)\pi/8$, the sign problem is most severe and finite-size scaling corrections are most pronounced; consequently, the local peak observed below the phase boundary in the temperature scan is a finite-size artifact, not a genuine new phase. While this cannot be rigorously verified by finite-size Monte Carlo data, it can be understood as follows: the positive and negative signs appear with almost equal probability, causing the average sign to approach zero as the system size increases. This indicates that the order in this region originates from a superposition of configurations with different signs, and cannot be a long-range order formed by a single configuration.

\subsection{Average sign and its derivatives in relation to phase transitions}
Guided by the complete phase diagram established here [Fig.~\ref{fig:phase_diagram}(e)], we now revisit our previous studies on the average sign~\cite{signGBW}, which were based on an incomplete phase diagram. Moreover, the WL algorithm employed here allows us to compute not only the average sign and its derivatives, but also the free energy and its derivatives, thus overcoming the limitation of previous studies that focused solely on the average sign. This enables a more thorough and systematic investigation of the relation between the average sign, its derivatives, and the phase transitions in the model.

For this purpose, we adopt the same two temperature paths used to verify the phase diagram. The first, $(K,\phi) = \beta(0.41222, 0.2)$, crosses SDA at a known genuine phase boundary. The second, $(K,\phi) = \beta(0.40788, 1.0)$, traverses three key points: the genuine boundary on SDB at $(0.40788, 1.0)$; the dashed (non-critical) segment of SDA at $(0.22269, 0.54624)$; and $(0.16017, \pi/8)$, where the cosine factor is extremal and the sign problem is most severe. Both paths are marked in Fig.~\ref{fig:phase_diagram}(c).

\begin{figure}[htb]
	\includegraphics[width=0.45\textwidth]{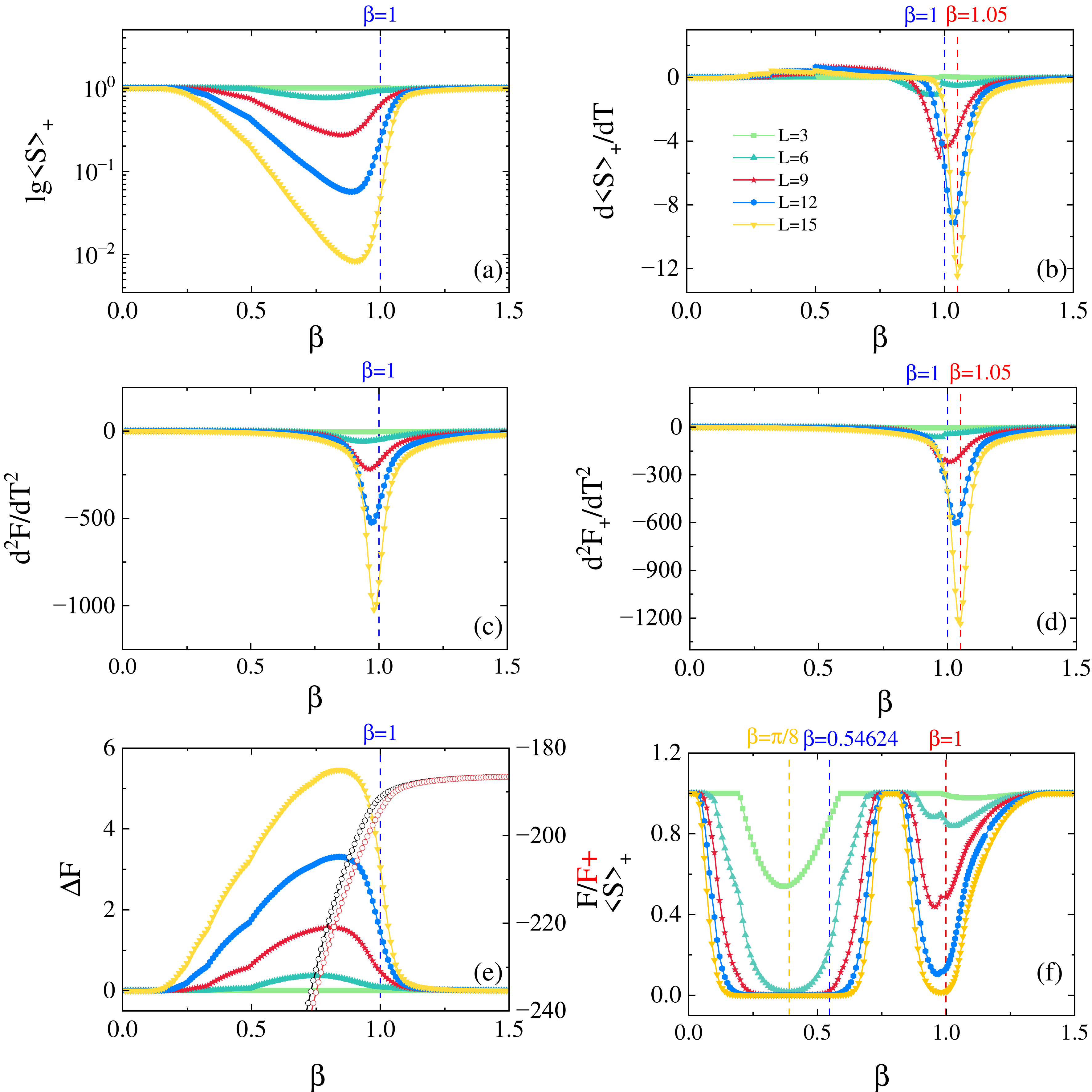}
	\caption{
    From WL simulations with system sizes $L=3$ to $15$, we study the average sign and its derivatives, as well as the free energy and its second derivative, as functions of $\beta$. Along the temperature path $(K,\phi)=\beta(0.41222,0.2)$: (a) The minimum of $\langle S\rangle_+$ converges to $\beta\simeq 1$, signaling the transition of $\mathcal{Z}$. (c) The minimum of $d^2F/dT^2$ converges to $\beta\simeq 1$, locating the transition of $\mathcal{Z}$. (e) The maximum of $\Delta F = F - F_+$ (left axis) at $\beta\simeq 1$ corresponds to the minimum of $\langle S\rangle_+$; the right axis shows $F$ and $F_+$, with $\mathcal{Z}_+$ transitioning at larger $\beta$. (b) The minimum of $d\langle S\rangle_+/dT$ occurs at $\beta\simeq 1.05$, shifted from the critical point of $\mathcal{Z}$. (d) The minimum of $d^2F_+/dT^2$ at $\beta\simeq 1.05$ matches that in (b), confirming that the derivative of the sign probes the transition of $\mathcal{Z}_+$. (f) Along the path $(K,\phi)=\beta(0.40788,1.0)$, the minimum of $\langle S\rangle_+$ at $\phi=\pi/8$ arises from a minimum of $\mathcal{Z}$ coinciding with a maximum of $\mathcal{Z}_+$, which is a periodic artifact of the cosine factor rather than a thermodynamic transition.
    %From WL simulations ($L=3$--$15$), the average sign $\langle S\rangle_+$, its temperature derivative, free-energy difference, and second derivatives of the free energy are shown as functions of $\beta$. Path 1: $(K,\phi)=\beta(0.41222,0.2)$ ((a)--(e), genuine transition at $\beta\simeq1$); Path 2: $(K,\phi)=\beta(0.40788,1.0)$ ((f)). (a) The minimum of $\langle S\rangle_+$ converges to $\beta\simeq1$, signaling the transition of $\mathcal{Z}$. (b) The minimum of $d\langle S\rangle_+/dT$ is at $\beta\simeq1.05$, shifted from $\mathcal{Z}$'s critical point. (c) The minimum of $d^2F/dT^2$ converges to $\beta\simeq1$, locating the transition of $\mathcal{Z}$. (d) The minimum of $d^2F_+/dT^2$ at $\beta\simeq1.05$ matches that in (b), confirming that the sign derivative probes the transition of $\mathcal{Z}_+$. (e) The maximum of $\Delta F=F-F_+$ (left axis) at $\beta\simeq1$ corresponds to the minimum of $\langle S\rangle_+$; right axis shows $F$ and $F_+$, with $\mathcal{Z}_+$ transitioning at larger $\beta$. (f) For Path 2, the second minimum of $\langle S\rangle_+$ at $\phi=\pi/8$ arises from a minimum of $\mathcal{Z}$ coinciding with a maximum of $\mathcal{Z}_+$, which is a periodic artifact of the cosine factor, not a thermodynamic transition.
    }
	\label{fig:WL-sign}
\end{figure}

Fig.~\ref{fig:WL-sign}(a)--(e) presents the results for the first temperature path $(K,\phi) = \beta(0.41222, 0.2)$. Fig.~\ref{fig:WL-sign} (a) shows the average sign $\langle S\rangle_+$ as a function of inverse temperature $\beta$. A clear minimum develops as the system size increases, and its position converges toward $\beta_c$, which has been identified as a genuine phase transition. Fig.~\ref{fig:WL-sign} (c) displays the second-order derivative of the free energy of the original system defined as:
\begin{equation}
\label{eq:F2}
\frac{d^2F}{dT^2} = -2 \frac{d\mathcal{Z}/dT}{\mathcal{Z}(T)}
- T \frac{d^2\mathcal{Z}/dT^2}{\mathcal{Z}(T)}
+ T \left( \frac{d\mathcal{Z}/dT}{\mathcal{Z}(T)} \right)^2,
\end{equation}
where $d\mathcal{Z}/dT$ and $d^2\mathcal{Z}/dT^2$ are given Eqs.(\ref{eq:app_Z1}) and (\ref{eq:app_Z2}) in the Appendix\ref{app:partition}. Its minimum also converges to the same $\beta_c$ with increasing system size. The coincidence of the minima in Fig.~\ref{fig:WL-sign} (a) and (c) indicates that, at this parameter point, the minimum of the average sign follows the genuine phase transition of the original model.

However, by definition, $\langle S\rangle_+ = \mathcal{Z}/\mathcal{Z}_+$ is the ratio of the partition functions of the original and reference systems~\cite{PhysRevB.110.125141}. This mixed nature is reflected in Fig.~\ref{fig:WL-sign} (e), which shows the free-energy difference $\Delta F = F - F_+ = -T\ln\langle S\rangle_+$, along with $F$ and $F_+$ themselves for the largest system size $L=15$. The maximum of $\Delta F$ coincides with the minimum of $\langle S\rangle_+$, confirming that the sign behavior encodes information from both $\mathcal{Z}$ and $\mathcal{Z}_+$. Notably, the inflection point of $F$ occurs at $\beta \simeq 1$, while that of $F_+$ occurs at a slightly larger $\beta$, indicating that the two systems undergo phase transitions at distinct temperatures. We observe that for $L=15$, $F_+$ begins to bend while $F$ remains nearly flat, suggesting that the derivative of the average sign is likely probing the transition of the reference system $\mathcal{Z}_+$. Compared with our previous work~\cite{signGBW}, where $\Delta F$ was obtained by exact enumeration up to $L=6$, the WL approach employed here allows us to extend the calculation to $L=15$, providing a more robust finite-size analysis.

To confirm the relation between the derivative of the average sign and the phase transition of the reference system, we examine $d\langle S\rangle_+/dT$ and the second-order derivative of the free energy of the reference system, $d^2F_+/dT^2$, shown in Fig.~\ref{fig:WL-sign}(b) and (d), respectively. The derivatives are given by:
\begin{equation}
\frac{d\langle S\rangle_+}{dT} = \frac{1}{\mathcal{Z}_+^2}\left(\frac{d\mathcal{Z}}{dT}\mathcal{Z}_+ - \mathcal{Z}\frac{d\mathcal{Z}_+}{dT}\right),
\end{equation}
and
\begin{equation}
\frac{d^2F_+}{dT^2} = -2 \frac{d\mathcal{Z}_+ /dT}{\mathcal{Z}_+(T)}
- T \frac{d^2\mathcal{Z}_+/dT^2}{\mathcal{Z}_+(T)}
+ T \left( \frac{d\mathcal{Z}_+/dT}{\mathcal{Z}_+(T)} \right)^2,
\end{equation}
where the explicit forms of $d\mathcal{Z}_+/dT$ and $d^2\mathcal{Z}_+/dT^2$ are given in Eq.(\ref{eq:app_Zp1}) and Eq.(\ref{eq:app_Zp2}) of the Appendix~\ref{app:partition}.
The minima of these two quantities converge to the same temperature, $\beta \simeq 1.05$, with increasing system size. This confirms that the inflection point of $F_+$ seen in Fig.~\ref{fig:WL-sign}(e) is indeed slightly above $\beta=1$, consistent with the observation that the two free energies probe distinct phase transitions. Thus, for the temperature path parameterized by $(K,\phi)=\beta(0.41222,0.2)$, the minimum of $\langle S\rangle_+$ signals the transition of the original model, while the minimum of its derivative signals the transition of the reference model.

The situation changes qualitatively for the second temperature path, parameterized by $(K,\phi) = \beta(0.40788, 1.0)$, as shown in Fig.~\ref{fig:WL-sign}(f). The average sign now exhibits two distinct minima as a function of $\beta$. The first minimum, located at $\beta \simeq 1$, corresponds to the genuine phase transition on the SDB line, consistent with the finite-size scaling analysis presented in Sec.~\ref{sec:results}~A. The second minimum appears at $\beta \simeq \pi/8$, where $K = 0.16017 < K_c = 0.32924$, placing this point outside the true critical region of the self-dual lines. This spurious minimum arises from the intrinsic periodicity of the cosine factor in the bundled weight, which reaches an extremum at $\phi = \pi/8$ as shown in Fig.~\ref{fig:phase_diagram}(b). At this point, the partition function $\mathcal{Z}$ of the original system exhibits a minimum while $\mathcal{Z}_+$ of the reference system shows a maximum, leading to a dip in their ratio $\langle S\rangle_+ = \mathcal{Z}/\mathcal{Z}_+$ that is purely a consequence of the weight structure rather than a thermodynamic singularity. This mechanism—where the minimum of $\langle S\rangle_+$ originates from a coincidence of a minimum of $\mathcal{Z}$ and a maximum of $\mathcal{Z}_+$—is not covered by the three general scenarios discussed in Ma et al.~\cite{PhysRevB.110.125141}, and thus provides a new concrete example of how the average sign can fail as a phase-transition probe.

Overall, these observations demonstrate that the minimum of the average sign can originate from either a genuine phase transition (first path) or a non-critical feature induced by the periodicity of the cosine factor (second path). In contrast, the minimum of the derivative of the average sign probes the phase transition of the reference system in the present GBW model, where $\mathcal{Z}_+$ undergoes a transition while $\mathcal{Z}$ remains relatively flat. More generally, as discussed by Ma et al.~\cite{PhysRevB.110.125141}, the derivative of the average sign can identify the phase transition of either the original or the reference system, depending on which of the two partition functions varies more rapidly. These findings clarify the capabilities and limitations of sign-based probes for detecting phase transitions in systems with sign problems.

\subsection{Exponential barrier in the Wang-Landau algorithm}

For brute-force reweighting, the origin of the exponential barrier is well understood: it stems from the exponential decay of the average sign with system size~\cite{Troyer2005, Loh1990}; when the decay is only algebraic, the barrier is absent. For WL, however, the situation is less clear. Its quantum generalization, the LLR algorithm, exhibits strong model dependence---polynomial in some models, exponential in others~\cite{LLR2014, LLR2020}. In this section, we examine whether WL can overcome the exponential barrier in the present model.

\begin{figure}[htb]
	\includegraphics[width=0.45\textwidth]
	{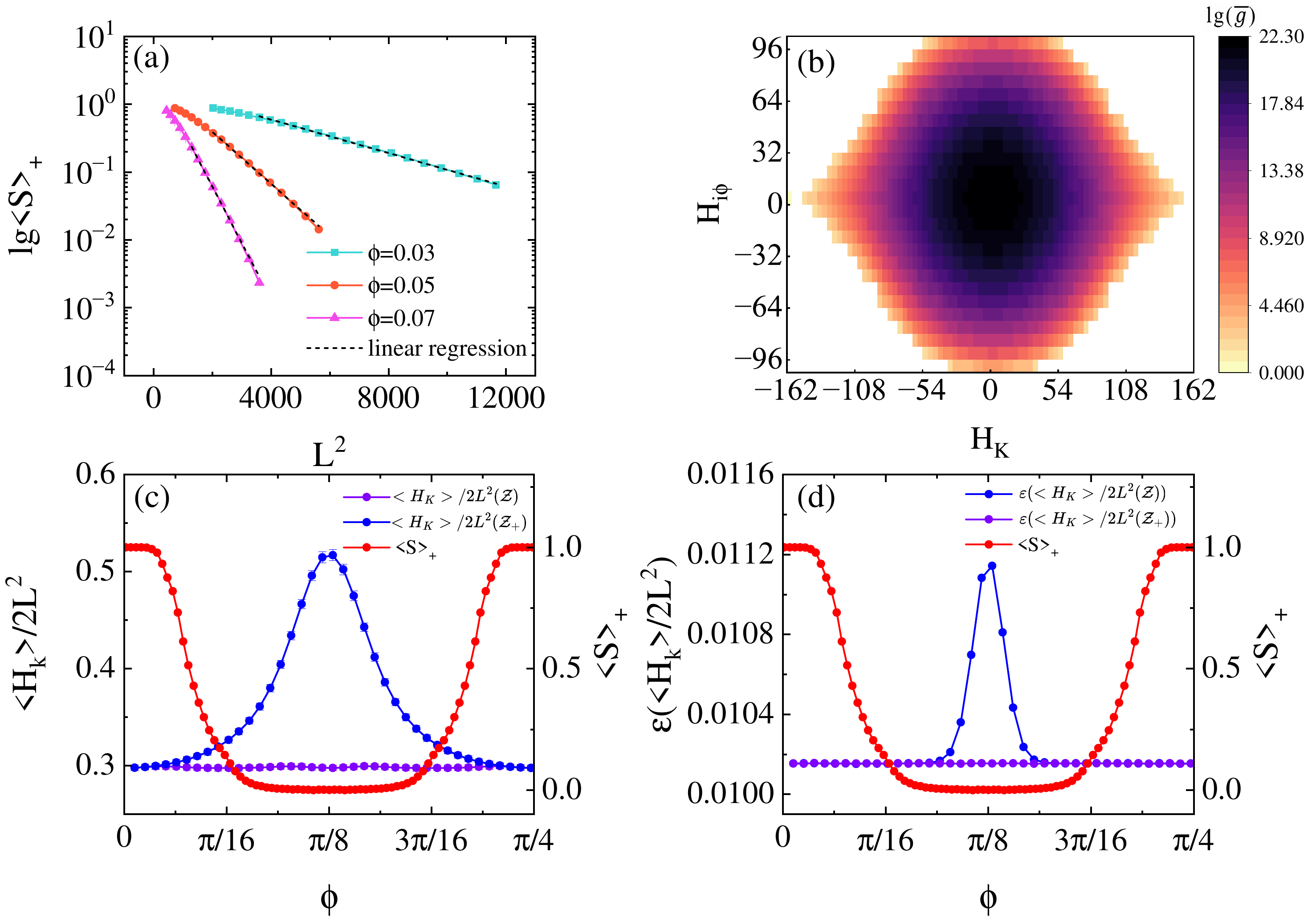}
	\caption{(a) Average sign at the critical point for $\phi=0.03$, $0.05$, $0.07$ on the self-dual line. $\lg\langle S\rangle_+$ decays linearly with $L^2$, giving $\langle S\rangle_+\propto e^{-\alpha L^2}$. (b) Logarithm of the average density of states $\lg \bar{g}(H_K,H_{i\phi})$ in the $(H_K,H_{i\phi})$ parameter space. (c) $\langle H_K\rangle/2L^2$ as a function of $\phi$ at $L=9$ for the original system $\mathcal{Z}$ and the reference system $\mathcal{Z}_+$, obtained from 100 independent WL runs. (d) Relative error $\varepsilon_{\rm WL}$ of $\langle H_K\rangle/2L^2$ (left axis) and $\langle S\rangle_+$ (right axis) as functions of $\phi$ at $L=9$.}
	\label{fig:error-WL}
\end{figure}

We first examine the behavior of the average sign in the GBW model. Fig.~\ref{fig:error-WL}(a) shows the average sign at the critical point for three representative values $\phi=0.03, 0.05, 0.07$ on the genuine phase boundary, obtained via Metropolis sampling on the reference system $\mathcal{Z}_+$. The data show that $\lg\langle S\rangle_+$ decays linearly with $L^2$, i.e., $\langle S\rangle_+ \propto e^{-\alpha L^2}$, with the points falling precisely on the linear fit as the system size increases. This exponential decay of the average sign with system volume is intrinsic to the model.

To understand the impact of the sign problem on the statistical error of the WL algorithm, we compare the partition function of the original system Eq.~(\ref{ZPFg}) with a reference system whose partition function is：
\begin{equation}
\mathcal{Z}_+(T) = \sum_{(H_K,H_{i\phi})} g(H_K,H_{i\phi}) \left|W(H_K, H_{i\phi})\right|.
\end{equation}
 Both share the same set of variables $(H_K, H_{i\phi})$ and the same density of states in this space. The only difference lies in the Boltzmann weight: $W(H_K,H_{i\phi})$ for the original system versus $\left|W(H_K,H_{i\phi})\right|$ for the reference system. Fig.~\ref{fig:error-WL}(b) displays the average density of states $\bar{g}(H_K, H_{i\phi})$ obtained from 100 independent WL simulations with different random seeds. We restrict the system size to $L=9$ in order to achieve higher precision and to resolve the details of the error evolution.

Using these 100 sets of density of states, we compute the statistical errors of the same physical observable in both the original system $\mathcal{Z}$ and the reference system $\mathcal{Z}_+$. Fig.~\ref{fig:error-WL}(c) shows $\langle H_K\rangle/2L^2$ as a function of $\phi$. The error bars for the original system are significantly larger than those for the reference system, and their variation with $\phi$ exhibits a trend opposite to that of the average sign. To exclude the influence of the magnitude of $\langle H_K\rangle/2L^2$ itself, we further calculate the relative error as a function of $\phi$.
The relative error in the original system is again substantially larger, with the same opposite trend. This confirms that the statistical error of the WL algorithm is severely affected by the sign problem.

To understand why the errors differ so markedly despite the identical density of states, we examine the observable formula explicitly. For the original system, any observable $O$ can be evaluated as:
\begin{equation}
\begin{aligned}
&\langle O \rangle
= \dfrac{\sum\limits_{(H_K,H_{i\phi})}\hspace{-0.2cm} O(H_K,H_{i\phi})g(H_K,H_{i\phi})W(H_K,H_{i\phi})}
        {\sum\limits_{(H_K,H_{i\phi})} g(H_K,H_{i\phi}) W(H_K,H_{i\phi})}= \\
& \dfrac{\sum\limits_{(H_K,H_{i\phi})}\hspace{-0.2cm} O(H_K,H_{i\phi})g(H_K,H_{i\phi})|W(H_K,H_{i\phi})|S(H_K,H_{i\phi})}
        {\sum\limits_{(H_K,H_{i\phi})} g(H_K,H_{i\phi})|W(H_K,H_{i\phi})|S(H_K,H_{i\phi})} \\
&= \dfrac{\langle O S \rangle_+}{\langle S \rangle_+},
\end{aligned}
\end{equation}
where $S(H_K,H_{i\phi}) = \operatorname{sgn}[W(H_K,H_{i\phi})]$, and the subscript $+$ denotes averaging with respect to the reference-system probability $P_+(H_K,H_{i\phi}) \propto g(H_K,H_{i\phi})|W(H_K,H_{i\phi})|$.

Since the WL algorithm computes the density of states in the $(H_K, H_{i\phi})$ space, which is common to both systems, the difference in errors observed in Fig.~\ref{fig:error-WL}(c) and (d) must originate from the sign problem. Specifically, the sign problem in the original system introduces severe cancellations that amplify the statistical fluctuations of the ratio $\langle O S \rangle_+ / \langle S \rangle_+$, whereas the absolute value $|W(H_K, H_{i\phi})|$ in the reference system removes this amplification. Thus, the WL error is not determined solely by the accuracy of the density-of-states estimation; rather, it is strongly influenced by the intrinsic sign problem of the target system.

Next, we quantitatively evaluate its error. It should be emphasized that the error of the WL algorithm arises from two combined sources: a systematic bias in the density-of-states estimation determined by the termination condition---namely, the reduction of the modification factor to a preset minimum $\ln f_{\min}$---and an amplification effect of the sign problem on the observables, which originates from the exponential decay of the denominator $\langle S\rangle_+$ in the reweighting formula. These two effects together determine the effective error of a single measurement:$\varepsilon_{\rm WL} \sim \varepsilon_{\rm DOS}/\langle S\rangle_+$
where $\varepsilon_{\rm DOS}$ is the error in the density-of-states estimation controlled by the termination condition, and $1/\langle S\rangle_+$ is the amplification factor arising from the sign problem. Within the WL framework, improving $\varepsilon_{\rm DOS}$ requires tightening the termination condition (e.g., setting a smaller $\ln f_{\min}$), which forces more iteration levels and more thorough flat sampling, so the computational cost per run scales as $T \propto \varepsilon_{\rm DOS}^{-\gamma}$ (with $\gamma>0$ being a model-dependent exponent)~\cite{PhysRevE.72.025701, PhysRevE.96.043307}. Since $\langle S\rangle_+ \propto e^{-\beta N}$, a single measurement already carries an error $\varepsilon_{\rm WL} \propto \varepsilon_{\rm DOS} e^{\beta N}$. To achieve the same observational precision as in the reference system (the sign-problem-free $\mathcal{Z}_+$ model), one can tighten the termination condition (i.e., reduce $\varepsilon_{\rm DOS}$) to compensate for the amplification effect of the sign problem, so the cost per run scales as:
\begin{equation}
\label{Eq:WLtime}
T_{WL} \propto 1/\langle S\rangle_+^{\gamma} \propto e^{\gamma \beta N},
\end{equation}
which grows exponentially with system size. Thus, even though the termination condition can be optimized to remove the impact of the sign problem on the precision, the exponential growth of the single-run cost still renders the WL algorithm impractical for large system sizes in the GBW model.

The above analysis demonstrates that the WL algorithm, although formally circumventing the sign problem, still cannot overcome the exponential barrier in the GBW model due to the combined effect of the finite precision of the density-of-states estimation and the amplification caused by the sign problem.

\section{Discussion and outlook}
\label{sec:conclusions}
Using self-duality analysis combined with unbiased Monte Carlo simulations, we have systematically investigated the periodicity-driven revision of the phase diagram of the generalized Baxter-Wu model with asymmetric complex couplings. Guided by the revised phase diagram, we have further explored the viability and limitations of sign-based phase-transition probes, as well as the question of whether density-of-states methods such as the WL algorithm can overcome the exponential barrier.

We have incorporated the $\pi/4$ periodicity of the partition function, encoded in the cosine factor of the bundled Boltzmann weight, into the self-duality analysis of the GBW model. This periodicity generates a discrete family of additional self-dual lines Eq.~(\ref{Eq:self_dual_line2}), which had been overlooked in previous transfer-matrix and Monte Carlo studies~\cite{blote2017,signGBW}. Guided by the complete set of self-dual candidates, our unbiased simulations, based on brute-force reweighting and the WL algorithm, establish that the self-dual lines at the partition-function minima $\phi_{\mathcal{Z}_{\min}}=(2n+1)\pi/8$ constitute a critical threshold. This threshold coincides with the intersection of the conventional family (SDA) and the newly identified family (SDB): the segments with $|K| \ge \frac{1}{2}\operatorname{arsinh}(\cos(\pi/4)) \approx 0.32924$ are genuine critical boundaries, whereas the segments with smaller $|K|$ are not, yielding the revised phase diagram in Fig.~\ref{fig:phase_diagram}(e). Data collapse of the squared order parameter $\langle m^2\rangle$ and the Binder ratio $Q$ using the four-state Potts exponents $\nu=2/3$ and $\eta=1/4$ confirms that the identified transitions belong to this universality class. In the vicinity of the partition-function minima, where the sign problem is most severe and the logarithmic corrections are strongest, the order parameter develops a local peak. Our simulations along the $\phi=\pi/8$ axis, presented in Sec.~\ref{sec:results} B, support that the peak of $\langle m^2\rangle$ is a finite-size artifact, rather than a genuine new phase: while the small-size Monte Carlo data alone cannot completely rule out a new phase at this point, the observed tendency of the average sign to approach zero with increasing system size strongly indicates that the signal originates from a superposition of configurations with opposite signs, rather than from long-range order formed by a single configuration. The periodicity-driven revision observed here is not necessarily a universal feature applicable to all models; whether such effects emerge depends on the specific microscopic details of the system. Nevertheless, it represents a factor that should not be overlooked when constructing phase diagrams from duality arguments. The self-dual line identifies candidates for phase boundaries, but rigorous confirmation ultimately requires unbiased numerical methods such as Monte Carlo simulations or transfer-matrix calculations.

The sign problem arises from the choice of representation, whereas genuine phase transitions in a physical system should not depend on such a choice. It is therefore somewhat unexpected that the average sign can serve as a probe of phase transitions. Indeed, as shown in Eq.~(\ref{Eq:average_sign1}), the average sign is determined by the ratio of the partition function of the original system, $\mathcal{Z}$, to that of the reference system, $\mathcal{Z}_+$. The representation does not affect the partition function $\mathcal{Z}$ of the original system at all, but it has a profound impact on the partition function $\mathcal{Z}_+$ of the reference system. If the free energy corresponding to the reference system $\mathcal{Z}_+$ becomes flat in the critical region of $\mathcal{Z}$, then detection of the phase transition based on both the average sign and its derivative becomes possible. Guided by the complete phase diagram, we re-examined the sign-based phase transition probes. Using the WL algorithm, we extended the calculation of $\langle S\rangle_+$ and $\Delta F$ from exact enumeration at $L=6$ to $L\le 15$, leading to two conclusions. First, minima of $\langle S\rangle_+$ can arise either from genuine transitions or from non-critical artifacts due to the cosine factor; thus, a minimum of $\langle S\rangle_+$ alone cannot identify a phase transition.  Second, extrema of $d\langle S\rangle_+/dT$ coincide with those of $d^2F_+/dT^2$, indicating that the derivative of the average sign probes the transition of the reference system $\mathcal{Z}_+$, not that of the original system $\mathcal{Z}$---consistent with the general discussion of Ma et al. that the derivative can indicate which of the two systems undergoes a transition~\cite{PhysRevB.110.125141}. Sign-based probes depend on the interplay between the original system and its reference counterpart. They may serve as useful cross-checks against direct thermodynamic observables, but they can hardly stand alone as an independent method for detecting phase transitions.

The error of the WL algorithm is determined by the precision of the density-of-states estimation $\varepsilon_{\rm DOS}$ (controlled by the termination condition) and the sign-amplification factor $1/\langle S\rangle_+$, with the single-measurement error given by $\varepsilon_{\rm WL} \sim \varepsilon_{\rm DOS}/\langle S\rangle_+$. To match the precision of the reference system, one can tighten the termination condition (i.e., reduce $\varepsilon_{\rm DOS}$) to offset the sign amplification, making the single-measurement error comparable to that of the reference.
Since $\langle S\rangle_+ \propto e^{-\beta N}$, achieving a given $\varepsilon_{\rm WL}$ requires $\varepsilon_{\rm DOS} \propto e^{-\beta N}$, so the per-run cost scales as $T_{\rm WL} \propto e^{\gamma \beta N}$, which grows exponentially with system size. Therefore, although the WL algorithm formally circumvents the sign problem, it does not truly circumvent the exponential barrier.
This is echoed by the LLR algorithm, which exhibits polynomial complexity $\propto V^p$ ($p\approx 2$) for the $\mathbb{Z}_3$ model at finite density~\cite{LLR2014}, while the exponential barrier reappears for the hexagonal Hubbard model~\cite{LLR2020}, confirming that the exponential barrier is model dependent, governed by the decay of the average sign rather than by the choice of the sampling method.

The periodicity-driven revision established here highlights the importance of accounting for such effects in constructing phase diagrams for complex-coupled models. The Hermitian transfer matrix of this model~\cite{blote2017,signGBW} implies unbroken PT symmetry and a real energy spectrum, suggesting that the sign problem can in principle be completely resolved~\cite{2010PTSP}. Based on this, we are currently conducting further exploration of sign-problem-free simulations to fully understand the phase transition mechanism. Although the WL algorithm cannot fully overcome the exponential barrier, it remains a valuable complement to brute-force reweighting methods for sign-problematic systems, owing to its advantages in mitigating critical slowing down, exploring phase transitions without prior order parameters, and covering the full parameter space in a single run.

\begin{acknowledgments}
We would like to thank Wenan Guo for the valuable discussions. Y. Liu acknowledges support from the National Natural Science Foundation of China under Grant Nos. 12305039 and 12574251, the Fundamental Research Funds for the Central Universities from the Beijing University of Posts and Telecommunications under Grant Nos. 2025JCTP08 and 2025AI4S01, as well as the Open Fund of the Key Laboratory of Multiscale Spin Physics (Ministry of Education), Beijing Normal University, under Grant No. SPIN2025K03.
\end{acknowledgments}
\appendix

\section{Partition function and its derivatives}
\label{app:partition}
Based on the Hamiltonian of the GBW model with asymmetric complex couplings [Eq.~(\ref{Eq:Ham})], the Boltzmann weight of a configuration $\Gamma\equiv(\sigma_1,\sigma_2,\ldots,\sigma_N)$ is given by:
\begin{equation}
\label{Eq:Boltzmann}
W(\Gamma, T) = e^{K H_K(\Gamma) + i\phi H_{i\phi}(\Gamma)},
\end{equation}
where the two integer-valued functionals $H_K$ and $H_{i\phi}$ are defined by
\begin{equation}
H_K(\Gamma) \equiv -\frac{\partial(\beta H)}{\partial K}
= \sum_{\triangle} \sigma_i\sigma_j\sigma_k + \sum_{\triangledown} \sigma_l\sigma_m\sigma_n,
\label{eq:app_HK}
\end{equation}
and
\begin{equation}
H_{i\phi}(\Gamma) \equiv -\frac{\partial(\beta H)}{\partial(i\phi)}
= \sum_{\triangle} \sigma_i\sigma_j\sigma_k - \sum_{\triangledown} \sigma_l\sigma_m\sigma_n
\label{eq:app_Hiph}
\end{equation}
which encode the sum and difference of up- and down-triangle three-spin product sums. The temperature dependence is carried by the reduced couplings $K = K_C/T$ and $\phi = \phi_C/T$.
In this paper, $(K_C, \phi_C)$ is defined as the point where a temperature scan crosses the actual phase boundary; for the green temperature path in Fig~\ref{fig:phase_diagram}(c),$(K_C,\phi_C)=(0.41222, 0.2)$.

Using the bundled-configuration scheme described in Sec.~\ref{sec:model} to circumvent the complexity arising from the complex-valued energy, the energy for the bundled configuration $\tilde{\Gamma}$ combining $\Gamma$ and $\Gamma'$ as shown in Fig.~\ref{fig:triangular_lattice}(b) can be expressed as a function of $H_K$ and $H_{i\phi}$:
\begin{equation}
\label{eq:E_bundled}
\begin{aligned}
\beta E(\tg)&=\frac{\beta[E(\Gamma) e^{-\beta E(\Gamma)} + E(\Gamma') e^{-\beta E(\Gamma')}]}{e^{-\beta E(\Gamma)} + e^{-\beta E(\Gamma')}}  \\
&= -KH_{K} + \phi H_{i\phi}\tan(\phi H_{i\phi}).
\end{aligned}
\end{equation}

Combining Eq.~(\ref{Eq:Wbindingconf}) and Eq.~(\ref{ZPFg}), the partition function is then also expressed as a function of  $H_K$ and $H_{i\phi}$:
\begin{equation}
\label{ZPFgg}
\mathcal{Z}(T) = \sum_{(H_{K},H_{i\phi})} g(H_{K},H_{i\phi}) e^{K H_{K}}\cos(\phi H_{i\phi}),
\end{equation}
where $g(H_K,H_{i\phi})$ is the density of states in the $(H_K,H_{i\phi})$ space.
Based on the analysis of single-spin flip processes, we determine the accessible values of $H_K$ and $H_{i\phi}$. The number of accessible values of $H_K$ is $n_K(L)=L^2-3$, and that of $H_{i\phi}$ is $n_{i\phi}(L)=\frac{L^2}{3}-(L\bmod 2)+1$. The total number of accessible $(H_K, H_{i\phi})$ combinations for each system size is listed in TABLE~\ref{tab:allowed_HK_Hiphi}.

\begin{table}[t]
\caption{ Numbers of distinct accessible values of $H_K$ and $H_{i\phi}$, denoted by $n_K$ and $n_{i\phi}$, respectively, and the total number $N(L)$ of accessible $(H_K,H_{i\phi})$ pairs. }
\label{tab:allowed_HK_Hiphi}
\centering
\renewcommand{\arraystretch}{1.15}
\begin{tabular*}{\columnwidth}{
@{\extracolsep{\fill}} c c c c @{}}
\hline\hline
$L$ & $n_K$ & $n_{i\phi}$ & $N(L)$ \\
\hline
$3$  & $6$   & $3$   & $10$ \\
$6$  & $33$  & $13$  & $229$ \\
$9$  & $78$  & $27$  & $1{,}310$ \\
$12$ & $141$ & $49$  & $4{,}333$ \\
$15$ & $222$ & $75$  & $10{,}718$ \\
$18$ & $321$ & $109$ & $22{,}625$ \\
\hline\hline
\end{tabular*}
\end{table}

Although the individual value sets of $H_K$ and $H_{i\phi}$ are known, not every pair in their Cartesian product is accessible because the two variables are subject to additional configurational constraints. For the six sizes listed in Table~\ref{tab:allowed_HK_Hiphi},
the number of accessible $(H_K,H_{i\phi})$ combinations is represented by the unified interpolation formula,
\begin{equation}
\label{eq:NHkHiphi}
\begin{split}
N(L)={}&\frac{2}{9}L^4-\frac{19}{6}L^2
+\frac{103}{6}L-\frac{169}{4} \\
&+(-1)^L\left(
\frac{2}{9}L^2-\frac{1}{6}L-\frac{51}{4}
\right).
\end{split}
\end{equation}
This expression gives $N(L)=10$, $229$, $1{,}310$, $4{,}333$, $10{,}718$, and $22{,}625$ for $L=3$, $6$, $9$, $12$, $15$, and $18$, respectively. It exactly interpolates these six determined values, although its validity as a general counting formula beyond $L=18$ has not yet been established.

Having characterized the space in which $g(H_K,H_{i\phi})$ resides, we next derive the partition-function derivatives required for the phase diagram analysis. The first and second derivative of the  partition function Eq.(\ref{ZPFgg}) with respect to $T$ are：

\begin{equation}
\frac{\partial\mathcal{Z}}{\partial K} = \sum_{(H_K, H_{i\phi})} g(H_K, H_{i\phi}) \, H_K \, e^{K H_K} \cos(\phi H_{i\phi})
\label{eq:app_Zp_derivK1}
\end{equation}
and
%The second derivative of the partition function with respect to $K$:
\begin{equation}
\frac{\partial^2\mathcal{Z}}{\partial K^2} = \sum_{(H_K, H_{i\phi})} g(H_K, H_{i\phi}) \, H_K^2 \, e^{K H_K} \cos(\phi H_{i\phi})
\label{eq:app_Zp_derivK2}
\end{equation}

The first and second derivative of the  partition function Eq.(\ref{ZPFgg}) with respect to $T$ are:
\begin{equation}
\begin{split}
\frac{d\mathcal{Z}}{dT} &= \sum_{(H_K, H_{i\phi})} g(H_K, H_{i\phi})
\frac{\partial}{\partial T}\Bigl[e^{K H_K}\cos(\phi H_{i\phi})\Bigr] \\
&= \sum_{(H_K, H_{i\phi})} g(H_K, H_{i\phi})\, e^{K H_K}
\Bigg[\frac{-K_C H_K}{T^2} \cos(\phi H_{i\phi})\\
&+ \frac{\phi_C H_{i\phi}}{T^2} \sin(\phi H_{i\phi})\Bigg], \\
\label{eq:app_Z1}
\end{split}
\end{equation}
and
\begin{equation}
\begin{split}
&\frac{d^2\mathcal{Z}}{dT^2} = \sum_{(H_K, H_{i\phi})} g(H_K, H_{i\phi}) \, e^{K H_K} \frac{1}{T^4}\\
&\Bigg[\left( K_C^2 H_K^2 + 2 K_C H_K T - \phi_C^2 H_{i\phi}^2 \right) \cos(\phi H_{i\phi})\\
&- 2 \phi_C H_{i\phi} (K_C H_K + T) \sin(\phi H_{i\phi})\Bigg]\\
\label{eq:app_Z2}
\end{split}
\end{equation}

Writing $|\cos(\phi H_{i\phi})| = \mathrm{sgn}(\cos(\phi H_{i\phi})) \cdot \cos(\phi H_{i\phi})$ and noting that $\mathrm{sgn}(\cos(\phi H_{i\phi}))$ is piecewise constant in $T$, we have:
\begin{equation}
\begin{split}
&\frac{\partial^n}{\partial T^n}\Bigl[e^{K H_K}|\cos(\phi H_{i\phi})|\Bigr]\\
&= \mathrm{sgn}\bigl(\cos(\phi H_{i\phi})\bigr)\,
\frac{\partial^n}{\partial T^n}\Bigl[e^{K H_K}\cos(\phi H_{i\phi})\Bigr].
\label{eq:app_Zp_deriv}
\end{split}
\end{equation}
Thus $\mathcal{Z}_+'(T)$ and $\mathcal{Z}_+''(T)$ follow immediately from Eqs.~(\ref{eq:app_Z1}) and (\ref{eq:app_Z2}) by inserting $\mathrm{sgn}(\cos(\phi H_{i\phi}))$ into each summand.
Consequently,
\begin{equation}
\begin{split}
\frac{d\mathcal{Z}_+}{dT} &= \sum_{(H_K, H_{i\phi})} g(H_K, H_{i\phi})\,
\mathrm{sgn}\bigl(\cos(\phi H_{i\phi})\bigr)\\
&\frac{\partial}{\partial T}\Bigl[e^{K H_K}\cos(\phi H_{i\phi})\Bigr], \label{eq:app_Zp1} \\
\end{split}
\end{equation}
\begin{equation}
\begin{split}
\frac{d^2\mathcal{Z}_+}{dT^2} &= \sum_{(H_K, H_{i\phi})} g(H_K, H_{i\phi})\,
\mathrm{sgn}\bigl(\cos(\phi H_{i\phi})\bigr)\\
&\frac{\partial^2}{\partial T^2}\Bigl[e^{K H_K}\cos(\phi H_{i\phi})\Bigr], \label{eq:app_Zp2}
\end{split}
\end{equation}
where the bracketed derivatives are exactly those evaluated in Eqs.~(\ref{eq:app_Z1}) and (\ref{eq:app_Z2}).
As Eq.~(\ref{eq:app_Zp_deriv}) shows, the derivatives of $\mathcal{Z}_+$ differ from those of $\mathcal{Z}$ solely by the insertion of $\mathrm{sgn}(\cos(\phi H_{i\phi}))$ into each summand. This is precisely the manifestation of the sign problem at the level of the partition function.
\clearpage
\bibliography{fassaad}

\end{document}